\documentclass[12pt]{article}
\newcommand{\bc}{\begin{center}}
\newcommand{\ec}{\end{center}}
\def\degr{\hbox{$^\circ$}}
\def\be{\begin{equation}}
\def\ee{\end{equation}}
\def\bea{\begin{eqnarray}}
\def\eea{\end{eqnarray}}
\def\l{\lambda}
\def\mkm{\mu{\rm m}}
\def\ga{\mathrel{\mathchoice {\vcenter{\offinterlineskip\halign{\hfil
$\displaystyle##$\hfil\cr>\cr\sim\cr}}}
{\vcenter{\offinterlineskip\halign{\hfil$\textstyle##$\hfil\cr
>\cr\sim\cr}}}
{\vcenter{\offinterlineskip\halign{\hfil$\scriptstyle##$\hfil\cr
>\cr\sim\cr}}}
{\vcenter{\offinterlineskip\halign{\hfil$\scriptscriptstyle##$\hfil\cr
>\cr\sim\cr}}}}}
\def\la{\mathrel{\mathchoice {\vcenter{\offinterlineskip\halign{\hfil
$\displaystyle##$\hfil\cr<\cr\sim\cr}}}
{\vcenter{\offinterlineskip\halign{\hfil$\textstyle##$\hfil\cr
<\cr\sim\cr}}}
{\vcenter{\offinterlineskip\halign{\hfil$\scriptstyle##$\hfil\cr
<\cr\sim\cr}}}
{\vcenter{\offinterlineskip\halign{\hfil$\scriptscriptstyle##$\hfil\cr
<\cr\sim\cr}}}}}
\def\noi{\noindent}
\def\bitem#1#2\par{\noindent\hangindent1.5\parindent\hangafter=1\rm#1
\rm#2\par\smallskip}
\usepackage{graphics}
\begin{document}
\thispagestyle{empty}
\phantom{.}
{\footnotesize\it Astronomy Letters, Vol. 26, N10, 2000.}
\bigskip
\bigskip
\hrule
\smallskip
\hrule
\bigskip
\bigskip

\bc{\large\bf
The Temperature of Nonspherical Circumstellar Dust Grains}\footnote{
To the memory of Gennadii Borisovich Sholomitskii,
an enthusiast for research in the field of infrared and
submillimeter astronomy.}\ec

\bigskip
\setcounter{footnote}{0}
\renewcommand{\thefootnote}{\fnsymbol{footnote}}
\centerline{\bf \copyright \, 2000 \,\,\,
N. V. Voshchinnikov\footnote{E-mail address for contacts:
nvv@dust.astro.spbu.ru.} and D. A. Semenov}

\begin{center}
{\it
Sobolev Astronomical Institute, St. Petersburg State University, \\
Bibliotechnaya pl. 2, St. Petersburg-Peterhof, 198504 Russia
}
\end{center}

\vskip 10pt

\centerline{\small Received January 28, 2000}

\begin{quote}\small
\noi {\bf Abstract}---The temperatures of prolate and oblate spheroidal dust grains
in the envelopes of stars of various spectral types are calculated.
Homogeneous particles with aspect ratios {\small $a/b \le 10$} composed of amorphous carbon,
iron, dirty ice, various silicates, and other materials are considered.
The temperatures of spherical and spheroidal particles were found to vary
similarly with particle size, distance to the star, and stellar temperature.
The temperature ratio {\small $T_{\rm d}({\rm spheroid})/T_{\rm d}({\rm sphere})$}  depends most strongly
on the grain chemical composition and shape. Spheroidal grains are generally
colder than spherical particles of the same volume; only iron spheroids
can be slightly hotter than iron spheres. At {\small $a/b \approx 2$}, the temperature
differences do not exceed 10\,\%. If {\small $a/b \ga 4$}, the temperatures can differ
by 30--40\,\%. For a fixed dust mass in the medium,
the fluxes at wavelengths {\small $\lambda \ga 100~\mkm$} are higher if the grains are
nonspherical, which gives over estimated dust masses from millimeter
observations. The effect of grain shape should also be taken into account
when modeling Galactic-dust emission properties,
which are calculated when searching for fluctuations of the cosmic
microwave background radiation in its Wien wing.

\noi{\it Keywords}: interstellar medium, circumstellar shells
\end{quote}

\bigskip
\centerline{\rm 1. INTRODUCTION }
\medskip

The observed infrared and submillimeter emission from interstellar clouds,
circumstellar envelopes, and galaxies is generally thermal emission of
dust heated by stellar radiation or shock waves. When infrared spectra of
these objects are computed, the dust temperature must be calculated.
This temperature is also used to determine the mass and thermal balance of
the matter in various objects and is important for the formation of molecules
on the grain surfaces.

Calculations of the interstellar dust temperature were initiated in
the 1940s [see van de Hulst (1949) for a discussion].
The equilibrium temperature of spherical dust grains is commonly
considered (see, e.g., Mathis {\it et al.} 1983).
However, it has been known for fifty years [since the
discovery of interstellar polarization by Hiltner (1949),
Hall (1949), and Dombrovskii (1949)] that there are
nonspherical aligned particles in the interstellar medium.

Nonspherical particles appear to be also present in
the circumstellar dust shells.
The variations in the
position angle of linear polarization with time and wavelength
observed in red giants provide circumstantial evidence for this
(Dyck and Jennings 1971; Shawl 1975).
Even after correction for the interstellar polarization,
the position-angle difference in the blue and in the red can reach $20^{\circ} - 60^{\circ}$.
This behavior is very difficult to explain in terms of the model of a
single star with a shell containing spherical particles alone.
If, alternatively, there are nonspherical grains in the shell,
then variations in the degree and direction of grain alignment may
result in the observable variations of the polarization angle.

The first attempt to take into account the effect of the shape
of interstellar grains on their temperature was made by
Greenberg and Shah (1971). These authors considered metallic and
dielectric Rayleigh spheroids and infinite ice cylinders
of 0.1-$\mu{\rm m}$ radius.
They concluded that nonspherical particles were approximately 10\,\%
colder than spheres, a result that entered the books on interstellar
dust (Whittet 1992).

Recently, Fogel and Leung (1998) have computed the infrared
radiation of fractal dust grains produced by two processes
of stochastic growth and composed of amorphous carbon and silicate.
They concluded that the temperature of nonspherical particles was
typically 10--20\,\% lower than that of spherical ones, which results in a
longward shift of the maximum of the object's radiation.
Fogel and Leung (1998) proposed to consider the fractal particle size
as a parameter of the grain shape, but they ignored the effects of alignment.

Here, we study in detail the dependence of the temperature of
spheroidal circumstellar dust grains on their shape and alignment.
We consider prolate and oblate particles of various sizes,
which are composed of a variety of absorbing and dielectric
materials and which lie at various distances from stars with
various temperatures. The nonsphericity effect of interstellar dust
grains on their temperature was discussed by Voshchinnikov {\it et al.} (1999).

\bigskip
\centerline{\rm 2. THE MODEL }
\medskip
\centerline{\it 2.1. The Radiation Field}
\medskip

Dust particles are present in the envelopes of late-type stars
(red giants and supergiants) and hotter stars, such as Herbig Ae/Be stars.
Dust grains are heated mainly by the absorption of stellar radiation.
As was pointed out by Lamy and Perrin (1997), in some cases,
it is important to take into account the star's true spectral energy
distribution as well. In addition, in the outer regions of optically
thick dust shells, the maximum in the energy distribution is
redshifted (see, e.g., Bagnulo {\it et al.} 1995). For simplicity,
however, we assume the energy distribution to be a blackbody one with
an effective temperature $T_{\star}$. Since we are going to compare the
temperatures of particles under the same conditions,
this assumption is of no fundamental importance.

In most cases, we take the stellar temperature to
be $T_{\star} = 2500$~K, typical of late-type giants and supergiants
(P$\acute{\rm e}$gouri$\acute{\rm e}$ 1987; Lorenz-Martins and Lef$\acute{\rm e}$vre 1995).
The effects of variations in $T_{\star}$ are discussed in subsection 4.5.

\medskip
\centerline{\it 2.2. Dust Particles }
\medskip

{\bf Chemical composition.}
Particles of amorphous carbon and amorphous silicates are most commonly
considered as the major sources of the infrared radiation observed
from carbon and oxygen stars, respectively. Evidence in support of these
materials follows both from theoretical models of dust formation
(Gail and Sedlmayr 1984) and from laboratory experiments
(J\"ager {\it et al.} 1994).

The specific type of silicate or carbon material in the circumstellar
medium is very difficult to determine. In addition, there is most
likely a mixture of various dust components in the shells simultaneously.
For example, bands of silicon carbide, sulfide silicates, and even
amorphous silicates were detected in the spectra of some carbon stars
(Baron {\it et al.} 1987; Goebel and Moseley; Little-Marenin 1986).
Several emission bands found in the spectra of carbon stars were
identified with crystalline silicates (Waters {\it et al.} 1999).
Finally, iron and oxide particles can apparently condense
in circumstellar envelopes irrespective of the C/O ratio.

\setcounter{footnote}{1}
\renewcommand{\thefootnote}{\arabic{footnote}}

In our modeling, we used the six materials that were previously
chosen by Il'in and Voshchinnikov (1998) when considering the effect of
radiation pressure on dust grains in the envelopes of late-type stars:
amorphous carbon, iron, and magnetite (Fe$_3$O$_4$) as examples of strongly
absorbing materials, as well as astronomical silicate (astrosil),
transparent glassy pyroxene, and artificial dirty silicate
(Ossenkopf {\it et al.} 1992; OHM silicate) as silicates of various types.
References to the papers from which we took the optical constants of these
materials can be found in Il'in and Voshchinnikov (1998).\footnote{Data on the refractive indices can also be extracted from
an electronic database of optical constants (Henning {\it et al.} 1999)
via Internet at http://www.astro.spbu.ru/JPDOC/entry.html.)}
This set of materials was extended to include carbon material (cellulose),
which was produced by pyrolysis at the temperature of 1000$^{\degr}$~C (cel1000;
J\"ager {\it et al.} 1998), and dirty ice, which was used in the classical
study by Greenberg and Shah (1971). In the latter case, the imaginary
part of the refractive index was chosen to be $k = 0.02$ in the wavelength
range 0.17--1.2 $\mkm$, as was done by Greenberg (1970, 1971).

{\bf Shape.}
The formation of only spherical particles in the envelopes of late-type stars
has been considered thus far (see, e.g., Draine 1981; Gail and Sedlmayr 1985;
Fadeyev 1987; Fleischer {\it et al.} 1992; Cadwell {\it et al.} 1994).
The theory of nucleation and growth of nonspherical particles is still at the
initial stage of its development. 
  
We assume the circumstellar dust grains to be prolate and oblate homogeneous
spheroids with aspect ratios $a/b$ ($a$ and $b$ are the spheroid semimajor and
semiminor axes, respectively). By varying $a/b$, we can model the particle
shape over a wide range: from spheres to needles and disks.

{\bf Size.}
Dust grains form and grow in the envelopes of late-type stars. They range in
size from tiny particles to particles with radii up to 1 $\mkm$ or more
[see Lafon and Berruyer (1991) for a discussion]. The upper limit of the
grain size distribution is uncertain and is the subject of debate.
However, it follows from model calculations that, in general, the particle
size in oxygen stars is larger than that in carbon ones
(Jura 1994, 1996; Bagnulo {\it et al.} 1995).

In order to compare the optical properties of particles of the same
volume but different shape, it is convenient to characterize the particle
size by the radius $r_{\rm V}$ of a sphere equal in volume to a spheroid. The spheroid
semimajor axis is related to $r_{\rm V}$ by
\be
a = r_{\rm V} \left( \frac{a}{b} \right)^{2/3}
\ee
for prolate spheroids and by
\be
a = r_{\rm V} \left( \frac{a}{b} \right)^{1/3}
\ee
for oblate spheroids. In our calculations, we considered particles
with $r_{\rm V} = 0.005 - 0.5\,\mkm$.

{\bf Structure.}
The dust grains growing in circumstellar shells can be fluffy or porous.
To model the effect of porosity, we used Bruggeman's rule (Bohren and
Huffman 1986) and obtained the mean effective dielectric function $\varepsilon_{\rm eff}$
of an aggregate composed of $n$ materials with dielectric functions
$\varepsilon_{i}$,
\be
\sum^n_{i=1} f_i \frac{\varepsilon_i - {\varepsilon_{\rm eff}}}
{\varepsilon_i + 2 {\varepsilon_{\rm eff}}}\ = 0,
\ee
where $f_{i}$ is the volume fraction occupied by the material of type $i$.
The temperature is calculated for compact particles with $\varepsilon_{\rm eff}$.
We considered spheroids composed of vacuum ($\varepsilon = 1$) and cellulose with
vacuum fractions from 0 to 0.9.

{\bf Orientation.}
Collisions of dust grains with atoms and molecules cause rapid grain
rotation with angular velocities $> 10^5$\,{\rm s}$^{-1}$.
Interstellar particles are
believed to rotate around the direction of maximum moment of inertia, and,
in general, the angular momentum is parallel to the magnetic field
(Spitzer 1981). Circumstellar particles can be aligned by anisotropic
radiation or gas fluxes (Dolginov {\it et al.} 1979). Radial grain motion
in the shells must apparently cause particle rotation in the planes
containing the radius vector. However, nonradial gas flows or the
helical circumstellar magnetic fields produced by stellar rotation
(see, e.g., Woitke {\it et al.} 1993) can also result in a different
grain alignment.

In our modeling, we considered two types of grain orientation: particles
randomly oriented in space (3D orientation) and in a plane (2D
orientation or complete rotational orientation). In the latter case, the
major axis of a rotating spheroid always lies in the same plane.
The angle $\Omega$ between the particle angular velocity and the wave
vector of the incident radiation is a model parameter
($0\degr \leq \Omega \leq 90\degr$).

\bigskip
\centerline{\rm 3. BASIC EQUATIONS  }
\medskip

Let us consider a dust grain at distance $R$  from a star of radius $R_{\star}$
and temperature $T_{\star}$. The stellar radiation is assumed to be unpolarized.
The equilibrium grain temperature $T_{\rm d}$ can be determined from Kirchoff's
law by solving the energy balance equation for the absorbed and
emitted energy (erg s$^{-1}$)
\be
W\,\int^\infty_0 \overline{C}_{\rm abs}(\l) \ \pi B_{\l}(T_{\star}) \,{\rm d}\l =
\int^\infty_0 \overline{C}_{\rm em}(\l) \ \pi B_{\l}(T_{\rm d}) \,{\rm d}\l\,, \label{eq1}
\ee
where  $\overline{C}_{\rm abs}(\l)$ and $\overline{C}_{\rm em}(\l)$
are the orientation-averaged absorption and
emission cross sections, $\pi B_{\l}(T)$ is the blackbody flux with temperature $T$
(erg cm$^{-2}$\,s$^{-1}\,\mkm^{-1}$), and $W = R_{\star}^2/R^2$ is the radiation
dilution factor.

For particles randomly oriented in space, the absorption cross sections
must be averaged over all orientations:
\be
\overline{C}^{\rm 3D}_{\rm abs} =
\int^{\pi /2}_0 \, \frac{1}{2} \,
\left[{Q}_{\rm abs}^{\rm TM}(m_\lambda,r_{\rm V},\l,a/b,\alpha) +
{Q}_{\rm abs}^{\rm TE}(m_\lambda,r_{\rm V},\l,a/b,\alpha)\right]
G(\alpha) \sin\alpha \,{\rm d}\alpha \,. \label{avc}
\ee
Here, $m_\lambda$ is the refractive index of the grain material
($m_\l = \varepsilon_\l^{1/2}$),
$\alpha$ is the angle between the spheroid rotation axis and the
wave vector ($0\degr \leq \alpha \leq 90\degr$), and $G$ is the geometric cross
section of the spheroid (the area of the particle shadow):
\be
G(\alpha) = \pi r_{\rm V}^2 \left (\frac{a}{b} \right ) ^{-2/3}
            \left[\left (\frac{a}{b} \right )^{2}\sin^2\alpha
            + \cos^2\alpha\right]^{1/2}
\label{Gp}
\ee
for prolate spheroids and
\be
G(\alpha) = \pi r_{\rm V}^2 \left (\frac{a}{b} \right ) ^{2/3}
            \left[\left (\frac{a}{b} \right )^{-2}\sin^2\alpha
            + \cos^2\alpha \right]^{1/2}\,.
\label{Go}
\ee
for oblate spheroids.

In the case of complete rotational orientation, the absorption cross
sections are averaged over all rotation angles $\phi$. For prolate spheroids,
this yields
\be
{\overline{C}}_{\rm abs}^{\rm \,2D} (\Omega) =
\frac{2}{\pi} \int_{0}^{\pi/2}\frac{1}{2}
\left[{Q}_{\rm abs}^{\rm TM}(m_\lambda,r_{\rm V},\l,a/b,\alpha)  +
 {Q}_{\rm abs}^{\rm TE}(m_\lambda,r_{\rm V},\l,a/b,\alpha) \right] G(\alpha)
       {\rm d}\phi,
\label{qabs2d}
\ee
where the angle $\alpha$ is related to $\Omega$ and $\phi$
($\cos\alpha = \sin\Omega \cos\phi$).
For oblate spheroids arbitrarily oriented in a plane, we have $\Omega = \alpha$ and
\be
{\overline{C}}_{\rm abs}^{\rm \,2D} (\Omega) =
\frac{1}{2}\left[{Q}_{\rm abs}^{\rm TM}(m_\lambda,r_{\rm V},\l,a/b,\Omega) +
{Q}_{\rm abs}^{\rm TE}(m_\lambda,r_{\rm V},\l,a/b,\Omega)\right] G(\Omega).
\label{qpr2do}
\ee
The energy emitted by a particle is proportional to its surface area.
The emission cross section can then be calculated as follows:
\be
\overline{C}_{\rm em} = S
\int^{\pi /2}_0 \, \frac{1}{2} \,
\left[{Q}_{\rm abs}^{\rm TM}(m_\lambda,r_{\rm V},\l,a/b,\alpha) +
{Q}_{\rm abs}^{\rm TE}(m_\lambda,r_{\rm V},\l,a/b,\alpha)\right]
 \sin\alpha \,{\rm d}\alpha \,, \label{ave}
\ee
where
\be
S = 2 \pi r_{\rm V}^2
            \left[\left (\frac{a}{b} \right )^{-2/3}
           + \left (\frac{a}{b} \right )^{1/3}\frac{\arcsin (e)}{e}
           \right]
\label{sp}
\ee
for prolate spheroids and
\be
S = 2 \pi r_{\rm V}^2
            \left[\left (\frac{a}{b} \right )^{2/3} +
            \left (\frac{a}{b} \right )^{-4/3}
            \frac{\ln [(1+e)/(1-e)]}{2 e}
           \right]
\label{so}
\ee
for oblate spheroids, $e=\sqrt{1-(a/b)^{-2}}$.

In Eqs. (5), (8), (9), and (10), the superscripts TM and TE refer to two
polarizations of the incident radiation (TM and TE modes).
The efficiency factors $Q_{\rm abs}^{\rm TM,TE}$ can be calculated using an exact solution to
the problem of diffraction of a plane electromagnetic wave by a
homogeneous spheroid by separation of variables method [see Voshchinnikov
and Farafonov (1993) for more detail). We used the benchmark results of
calculations of the efficiency factors from Voshchinnikov {\it et al.} (2000)
to thoroughly test the computational program.

\bigskip
\centerline{\rm 4. RESULTS AND DISCUSSION}
\medskip

In this section, we discuss the dependence of the spheroidal-grain temperature
on model parameters: the particle chemical composition, size, shape, and
orientation, as well as the stellar temperature and distance to the star.
We also consider the effects of porosity and partial polarization of the
incident radiation on the particle temperature. Since the main goal of our
study is to analyze the effects of change in the particle shape,
we compare the temperature ratio of spheroidal and spherical particles of
the same volume.

\medskip
\centerline{\it
4.1. A Graphical Method of Temperature Determination and a Blackbody Model
}
\medskip

This method, proposed by Greenberg (1970, 1971), allows the causes of
temperature differences between dust grains with different
characteristics to be clearly established if the particles
are assumed to be in an isotropic radiation field.
Their temperature can then be determined by solving the following
equation of thermal balance:
\be
W\,\int^\infty_0 \overline{C}_{\rm abs}(\l) \ 4 \pi B_{\l}(T_{\star}) \,{\rm d}\l =
\int^\infty_0 \overline{C}_{\rm em}(\l) \ \pi B_{\l}(T_{\rm d}) \,{\rm d}\l\,. \label{eq1a}
\ee
If the spheroidal particles are randomly oriented in space
(3D orientation), then numerical estimates show that
\be
\overline{C}^{\rm 3D}_{\rm abs} \approx \frac{\overline{C}_{\rm em}}{4}
\ee
The integrals on the left- and right-hand sides of Eq. (13) then depend
on the temperature $T$ alone (for given particle chemical composition,
size, and shape).

Figure 1 shows the power for the absorbed (emitted) radiation (in erg s$^{-1}$)
for spheres and spheroids of the same volume ($r_{\rm V} = 0.01\,\mkm$)
composed of cellulose (cel1000). The method of temperature determination
is indicated by arrows: from the point on the curve corresponding
to the stellar temperature ($T_{\star}$ = 2500 K), we drop a perpendicular whose
length is determined by the radiation dilution factor and then find the
point of intersection of the horizontal segment with the same curve
for the power. This point gives the dust grain temperature $T_{\rm d}$ determined
from the equation of thermal balance (13).

\begin{figure}\bc
\resizebox{8.5cm}{!}{\includegraphics{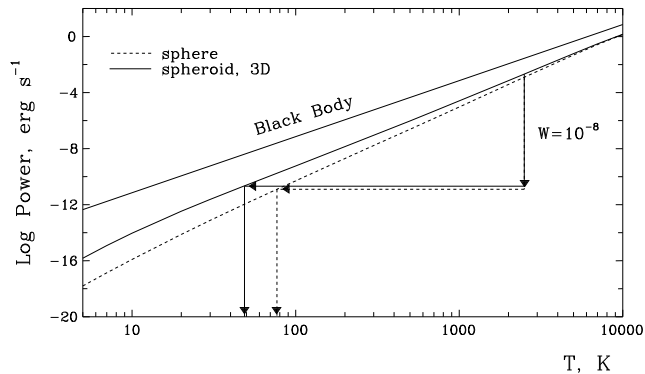}}
\caption[]{
The power absorbed (emitted) by dust particles
with $r_{\rm V}=0.01\,\mkm$ in an isotropic radiation field.
The straight line corresponds to a blackbody.
The curves refer to spherical and prolate spheroidal
($a/b=10$, 3D orientation) cellulose (cel1000) particles.
A graphical method of determining the grain temperature
is shown for $T_{\star} = 2500$~K and W$ = 10^{-8}$. In this case,
the temperatures of spherical and spheroidal particles
and a blackbody are respectively,
$T_{\rm d}({\rm sphere})=76.8$~K,
$T^{\rm 3D}_{\rm d}=48.7$~K, and
$T_{\rm d}^{\rm BB}= 2500 (10^{-8})^{1/4}= 25$~K.}
\label{f1}
\ec\end{figure}

It follows from Fig. 1 that the different temperatures of spheres and
spheroids result from different particle emissivities at low $T$. The
differences are most noticeable for cellulose. For other materials,
the pattern of dependence is preserved: the emissivity of spheroidal
particles at low temperatures is larger than that for spheres and
closer to the blackbody one (see Fig. 2). Only iron particles
constitute an exception: spheres composed of them at $T \la 30 \,{\rm K}$
emit more energy than spheroids with $a/b \simeq 2-6$ but less energy
than spheroids with $a/b \ga 8$ (Fig. 3).
\begin{figure*}\bc
\resizebox{12.0cm}{!}{\includegraphics{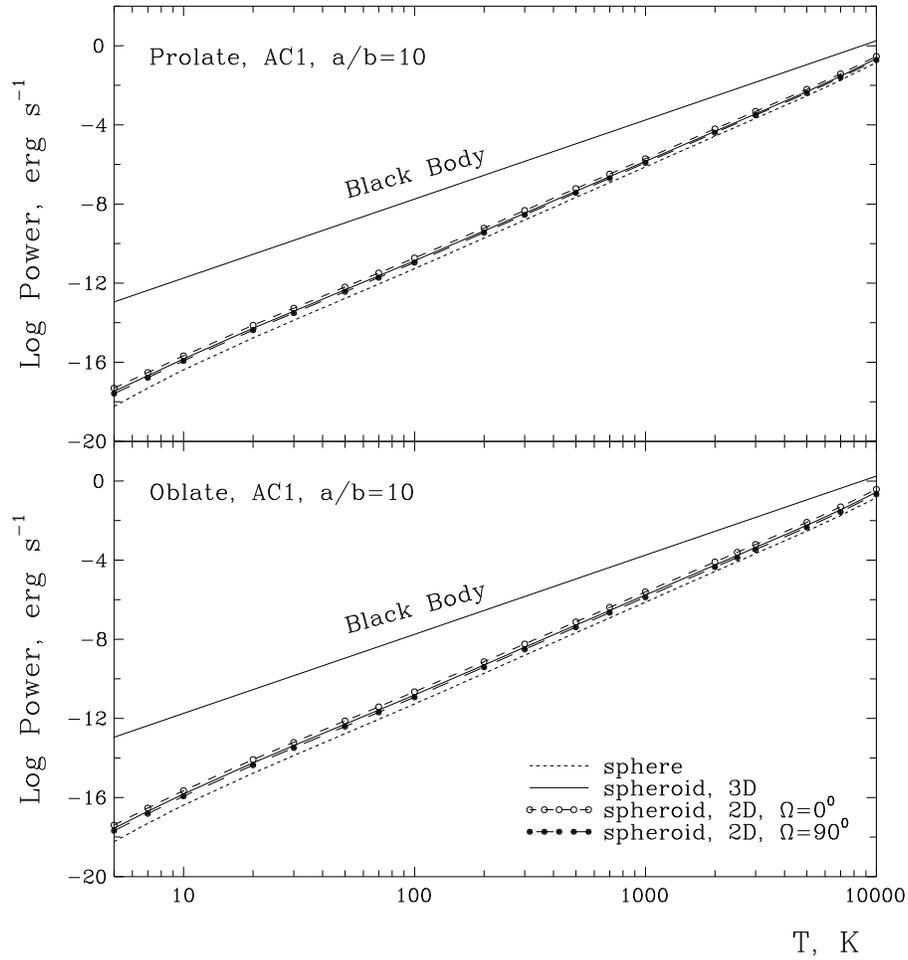}}
\caption[]{
The power absorbed by dust particles with $r_{\rm V} = 0.01\,\mkm$
in an anisotropic radiation field. The curves refer to spherical,
prolate, and oblate spheroidal amorphous carbon (AC1) particles with $a/b=10$.
The straight line corresponds to a blackbody.
}
\ec\end{figure*}
\begin{figure}\bc
\resizebox{8.5cm}{!}{\includegraphics{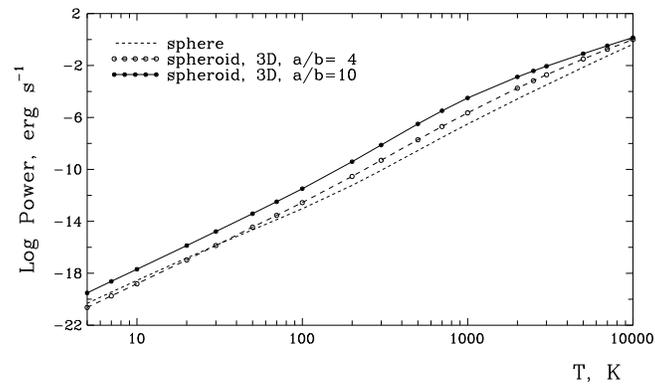}}
\caption[]{
The power absorbed by iron dust particles with $r_{\rm V}=0.01\,\mkm$
in an anisotropic radiation field.
}
\ec\end{figure}

The straight lines in the upper parts of Figs. 2 and 3 refer to a
blackbody, for which the emissivity is proportional to $T^4$, and
the absorption efficiency factors are equal to unity,
\be
{Q}_{\rm abs}^{\rm TM} = {Q}_{\rm abs}^{\rm TE} = 1 \,. \label{bb1}
\ee
Using Eq.~(15) and Stefan--Boltzmann's law, instead of Eq.~(13)
we obtain for the 3D orientation
\be
4 W \overline{G}^{\rm 3D} \sigma T_{\star}^{4} =
S  \sigma T_{\rm d}^{4} \,.
\ee
Considering that the mean projection area of any convex figure is a quarter
of its surface area (see, e.g., Hildebrand 1983),\footnote{Eq.~(17) for prolate
and oblate spheroids can be proved by directly integrating Eqs. (6)
and (7).}
\be
\overline{G}^{\rm 3D} = \frac{S}{4} \,, \label{bb2}
\ee
we have for spheres and spheroids randomly oriented in space
\be
 T_{\rm d}^{\rm BB} =  T_{\star} W^{1/4} \,.
\ee
For the 2D orientation, the blackbody temperature is
\be
 T_{\rm d}^{\rm BB} =  T_{\star}
 \left( \frac{4 W \overline{G}^{\rm 2D}(\Omega)}{S} \right )^{1/4} \,.
\ee
Expressions for the mean cross sections in the case of complete
rotational orientation were derived by Il'in and Voshchinnikov (1998): 
\be
\overline{G}^{\rm 2D}(\Omega) = \pi r_{\rm V}^2 \frac{2}{\pi}
E \left ( \left[1 - \left (\frac{a}{b} \right )^{-2} \right]
\sin^2\Omega \right ) \left ( \frac{a}{b} \right ) ^{1/3} \label{g2p}
\ee
for prolate spheroids and 
\be
\overline{G}^{\rm 2D}(\Omega) = {G}(\Omega) \label{g2o}
\ee
for oblate spheroids. In Eq.~(20), the complete elliptic integral of
the second kind is denoted by $E(m)$.
Based on Eqs.~(20) and (21), we can easily determine the
blackbody-temperature ratios for the two extreme cases of 2D
orientation ($\Omega = 0^{\circ}$ ¨ $\Omega = 90^{\circ}$):
\be
\frac{T_{\rm d}^{\rm BB}(\Omega = 0^{\circ})}
{T_{\rm d}^{\rm BB}(\Omega = 90^{\circ})} =
\left ( \frac{\overline{G}^{\rm 2D}(\Omega = 0^{\circ})}
{\overline{G}^{\rm 2D}(\Omega = 90^{\circ})}
\right ) ^{1/4} =
\left (\frac{\pi}{2 E \left ( 1 - \left (\frac{a}{b} \right )^{-2} \right )}
\right ) ^{1/4} \label{bb3}
\ee
for prolate spheroids and
\be
\frac{T_{\rm d}^{\rm BB}(\Omega = 0^{\circ})}
{T_{\rm d}^{\rm BB}(\Omega = 90^{\circ})} =
\left (\frac{a}{b} \right )^{1/4} \label{bb4}
\ee
for oblate spheroids. We can also calculate the temperature ratio of a
spheroid for the 2D orientation and a sphere (or a spheroid for the
3D orientation):
\be
\frac{T_{\rm d}^{\rm BB}(\Omega)}
{T_{\rm d}^{\rm BB}(\rm sphere, 3D)} =
\left (\frac{4 \overline{G}^{\rm 2D}(\Omega)}{S} \right )^{1/4} \,.
\label{bb5}
\ee

Note that Eqs.~(22)--(24) allow the geometric effects on the
temperature of nonspherical dust grains to be estimated.
For example,
$T_{\rm d}^{\rm BB}(\Omega = 0^{\circ}) / T_{\rm d}^{\rm BB}(\rm sphere) =
1.06$ (1.18) for $a/b = 10$
and
$T_{\rm d}^{\rm BB}(\Omega = 90^{\circ}) / T_{\rm d}^{\rm BB}(\rm sphere) =
0.95$ (0.66) for prolate (oblate) spheroids.

\medskip
\centerline{\it
4.2. Dependence on the Distance from the Star
}
\medskip

The temperature of particles of any shape decreases with distance from
the heating source (star). However, the temperature ratio of nonspherical
and spherical particles varies over a narrow range. As follows from Fig. 4,
which shows the results of our calculations for amorphous carbon particles,
this ratio decreases approximately by 5\,\% as the distance to the star
increases from 10 to $10^5~R_{\star}$. In this case, the sphere temperature
decreases from 780.6 to 20.8 K. Given the insignificant variation
of $T_{\rm d}$(spheroid)/$T_{\rm d}$(sphere) with $R$, below we consider the results
only for $R = 10^4 R_{\star} \, (W = 10^{-8})$.
\begin{figure}\bc
\resizebox{8.5cm}{!}{\includegraphics{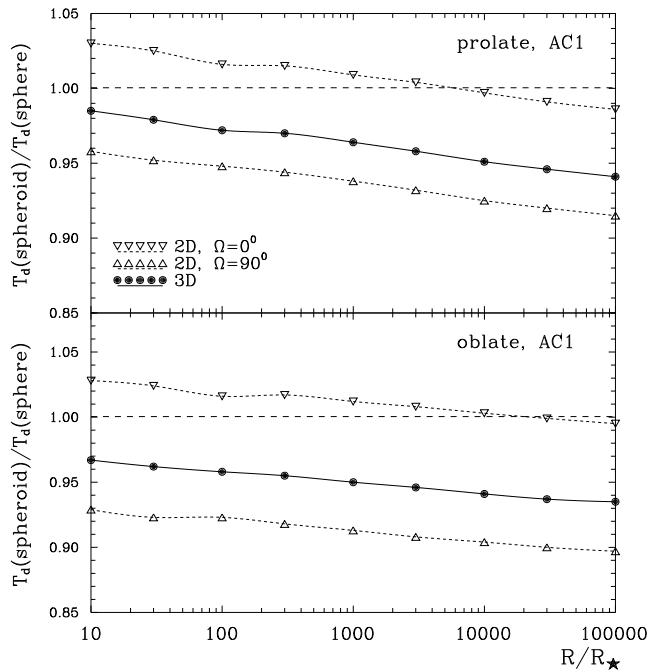}}
\caption[]{
The temperature ratio of prolate and oblate spheroidal
($a/b = 4$) and spherical amorphous carbon (AC1) grains with $r_{\rm V}=0.01\,\mkm$
at various distances from a star with $T_{\star}$ = 2500 K.
The sphere temperature is
$T_{\rm d}({\rm sphere})=$ 780.6~{\rm K}, 318.0~{\rm K},
130.8~{\rm K}, 52.3~{\rm K} and 20.8~{\rm K} for
$R = 10, 10^2, 10^3, 10^4$ and   $10^5~R_{\star}$, respectively.
}
\ec\end{figure}

\medskip
\centerline{\it
4.3. Dependence on the Grain Size
}
\medskip

The temperature variations of carbon and silicate dust grains with
particle radius are given in Tables 1 and 2. As $r_{\rm V}$ increases, the particle
temperature slightly rises and then begins to fall, as was also noted
previously (Greenberg 1970). The temperature of spheroidal particles is
always slightly lower than that of spherical ones; the difference
can reach 10\,\% at $a/b = 4$. The temperature difference between spheres and
spheroids is largest for the 2D orientation and $\Omega=90^{\circ}$.
Note also that, since $T_{\rm d}$ for spheres and spheroids vary with $r_{\rm V}$ similarly,
the temperature ratio is almost independent of $r_{\rm V}$.

\begin{table}
\bc
\caption[]{
The temperatures (in K) of spherical and spheroidal ($a/b=4$)
amorphous carbon grains; {\rm $T_{\star}= 2500\,$K}, $R = 10\,000\,R_{\star}$}
\smallskip
\begin{tabular}{cccccccccc}
\noalign{\smallskip}
\hline
\noalign{\smallskip}
$r_{\rm V}$,\,$\mkm$
& Sphere && \multicolumn{3}{c} Prolate spheroid
&& \multicolumn{3}{c} Oblate spheroid  \\
\noalign{\smallskip}
\cline{4-6} \cline{8-10}
\noalign{\smallskip}
&&&{\rm 2D, $\Omega=0^{\circ}$} & {\rm 2D, $\Omega=90^{\circ}$} & {\rm 3D} &&
{\rm 2D, $\Omega=0^{\circ}$} & {\rm 2D, $\Omega=90^{\circ}$} & {\rm 3D} \\
\noalign{\smallskip}
\hline
\noalign{\smallskip}
 0.005 & 52.3 &&  52.1 &  48.4 &  49.7 &&  52.4 & 47.2 & 49.2 \\
 0.010 & 52.3 &&  52.2 &  48.4 &  49.8 &&  52.4 & 47.3 & 49.2 \\
 0.020 & 52.4 &&  52.3 &  48.5 &  49.9 &&  52.6 & 47.4 & 49.3 \\
 0.030 & 52.6 &&  52.5 &  48.7 &  50.1 &&  52.8 & 47.6 & 49.5 \\
 0.050 & 53.3 &&  53.2 &  49.2 &  50.6 &&  53.3 & 48.2 & 50.1 \\
 0.100 & 56.0 &&  54.9 &  51.0 &  52.4 &&  54.8 & 50.4 & 52.0 \\
 0.200 & 61.1 &&  55.5 &  54.4 &  54.7 &&  54.6 & 54.3 & 54.3 \\
 0.300 & 62.3 &&  55.0 &  55.7 &  55.4 &&  53.6 & 55.4 & 54.6 \\
 0.500 & 60.2 &&  53.5 &  54.3 &  54.0 &&  51.8 & 53.6 & 52.9 \\
\noalign{ \smallskip}
\hline
\end{tabular}
\ec
\label{tab1}
\end{table}

\begin{table}
\bc
\caption[]{
The temperatures (in K) of spherical and spheroidal ($a/b=4$)
astronomical silicate grains $T_{\star}= 2500\,$K,
$R = 10\,000\,R_{\star}$}
\smallskip
\begin{tabular}{cccccccccc}
\noalign{\smallskip}
\hline
\noalign{\smallskip}
$r_{\rm V}$,\,$\mkm$
& Sphere && \multicolumn{3}{c} Prolate spheroid
&& \multicolumn{3}{c} Oblate spheroid  \\
\noalign{\smallskip}
\cline{4-6} \cline{8-10}
\noalign{\smallskip}
&&&{\rm 2D, $\Omega=0^{\circ}$} & {\rm 2D, $\Omega=90^{\circ}$} & {\rm 3D} &&
{\rm 2D, $\Omega=0^{\circ}$} & {\rm 2D, $\Omega=90^{\circ}$} & {\rm 3D} \\
\noalign{\smallskip}
\hline
\noalign{\smallskip}
 0.005 & 34.9 && 31.9  & 30.6  & 31.2  && 32.6 & 30.5 & 31.3  \\
 0.010 & 34.9 && 31.9  & 30.6  & 31.1  && 32.7 & 30.5 & 31.3  \\
 0.020 & 34.9 && 31.9  & 30.7  & 31.1  && 32.7 & 30.6 & 31.4  \\
 0.030 & 35.0 && 32.0  & 30.7  & 31.2  && 32.8 & 30.6 & 31.4  \\
 0.050 & 35.2 && 32.2  & 30.9  & 31.4  && 32.9 & 30.8 & 31.6  \\
 0.100 & 36.0 && 32.8  & 31.5  & 32.0  && 33.5 & 31.5 & 32.2  \\
 0.200 & 38.2 && 33.5  & 33.0  & 33.1  && 33.8 & 33.3 & 33.4  \\
 0.300 & 39.6 && 33.7  & 34.4  & 34.0  && 33.6 & 34.7 & 34.2  \\
 0.500 & 40.5 && 34.1  & 35.4  & 34.9  && 33.4 & 35.9 & 34.8  \\
\noalign{ \smallskip}
\hline
\end{tabular}
\ec
\end{table}

\medskip
\centerline{\it
4.4. Dependence on the Grain Shape and Chemical Composition
}
\medskip

The effects of the chemical composition and shape of dust grains on
their temperature are most significant compared to the variations of
other model parameters. These effects are illustrated in Figs. 5 and 6,
which show the temperature ratios of spheroids and spheres for particles
with metallic and dielectric properties, respectively.
The variations of $T_{\rm d}$(spheroid)/$T_{\rm d}$(sphere) have the following
general tendency: this ratio decreases with increasing $a/b$ and
absorptive material properties, with the effect
increasing. Besides, the following inequality holds for particles
of the same volume:
\be
T_{\rm d}({\rm sphere}) >
T^{\rm 2D}_{\rm d}(\Omega=0^{\circ}) >
T^{\rm 3D}_{\rm d} >
T^{\rm 2D}_{\rm d}(\Omega=90^{\circ}) \,.
\ee
\begin{figure}\bc
\resizebox{10.0cm}{!}{\includegraphics{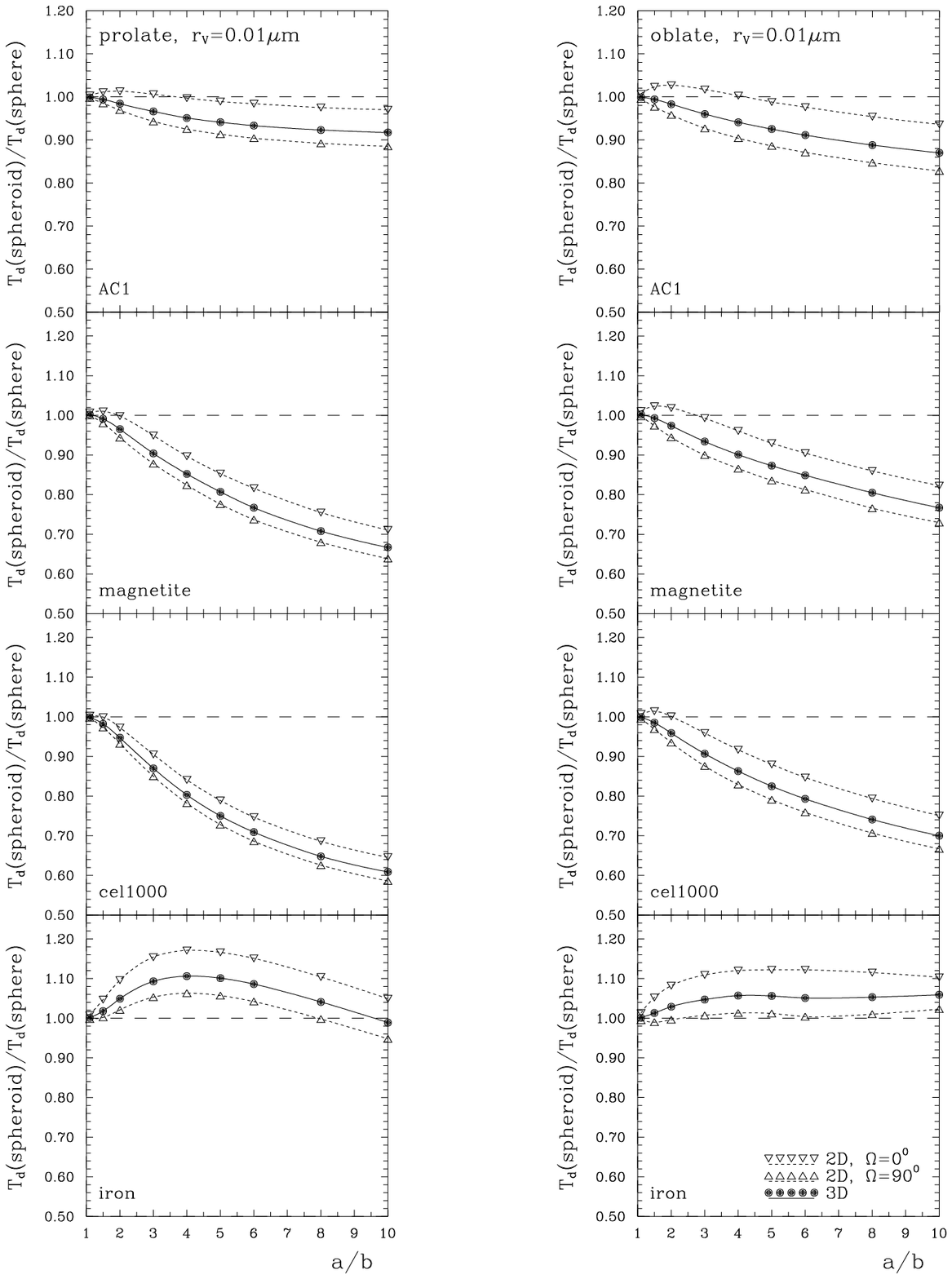}}
\caption[]{
The temperature ratio of prolate and oblate spheroidal and
spherical grains composed of various materials with metallic properties;
$T_{\star}$ = 2500 K, $r_{\rm V}=0.01\,\mkm$, and $R = 10^4~R_{\star}$.
The sphere temperature is
$T_{\rm d}({\rm sphere})=$ 52.3~{\rm K}, 63.0~{\rm K}, 59.2~{\rm K},
and 120.0~{\rm K}
for particles of amorphous carbon, magnetite, cellulose, and iron, respectively.
}
\ec\end{figure}
\begin{figure}\bc
\resizebox{10.0cm}{!}{\includegraphics{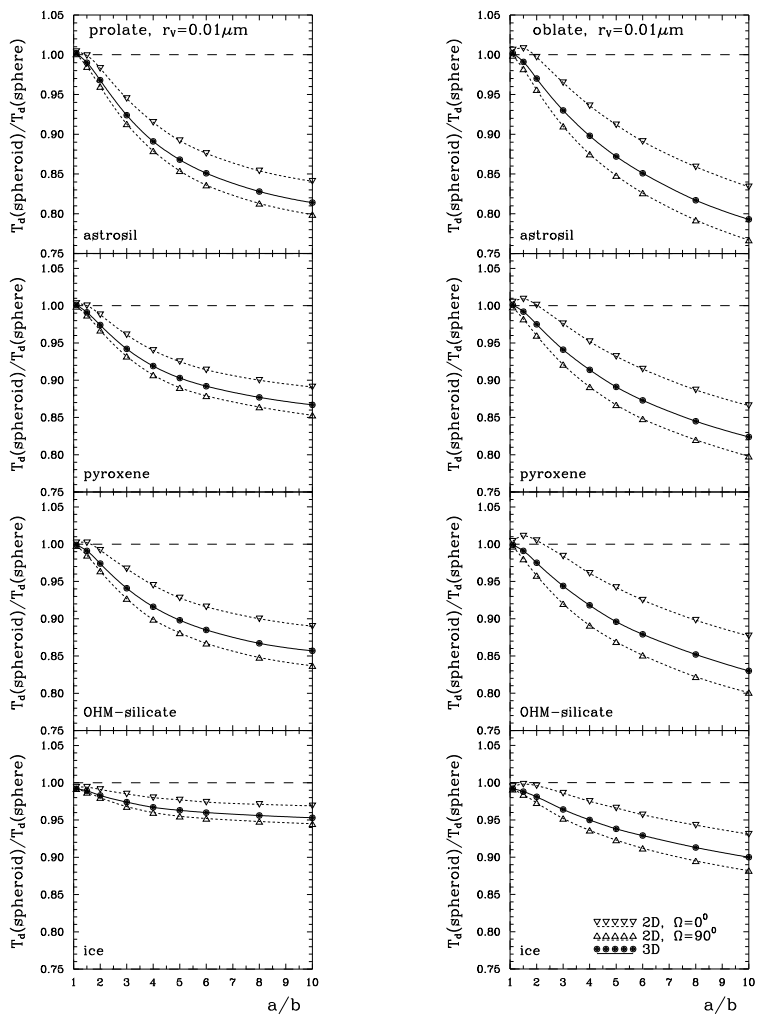}}
\caption[]{
The same as Fig. 5 for particles with dielectric properties.
The sphere temperature is
$T_{\rm d}({\rm sphere})=$ 34.9~{\rm K}, 24.0~{\rm K}, 42.4~{\rm K},
and 33.7~{\rm K} for
particles of astrosil, pyroxene, OHM silicate, and dirty ice, respectively.
}
\ec\end{figure}

The difference between prolate and oblate particles, which is noticeable
in Figs. 5 and 6, is determined by different optical properties of the
particles [see Voshchinnikov and Farafonov (1993) for more detail].
If we restrict ourselves to $a/b \la 4$, then it is easy to see that the
dielectric and metallic spheroids are, respectively, 10\,\% and 20\,\%
colder than the corresponding spheres. The temperatures of prolate and
oblate particles can differ by 30--40\,\% from the temperature of
spheres (Fig. 5). If we further increase the aspect ratio $a/b$,
then the temperature gradually ceases to drop. For example, for
prolate cel1000 particles,
$T^{\rm 3D}_{\rm d}/T_{\rm d}({\rm sphere}) =$ 0.61, 0.53, 0.50, and
0.50 for $a/b =$ 10, 20, 50, and 100, respectively.

The temperature behavior for iron particles is peculiar:
for $a/b \la 8$, the spheroids are 10--20\,\% hotter than the spheres.
The reason for this is seen from the behavior of the particle
emissivity shown in Fig. 3.

\medskip
\centerline{\it
4.5. Dependence of the Stellar Temperature
}
\medskip

Nonspherical particles may be present in the shells of stars of various
spectral types and in the interplanetary medium. The effects of
variations in the stellar temperature are shown in Fig. 7 for
amorphous carbon particles with $r_{\rm V}=0.01\,\mkm$.
Low effective temperatures (up to 500 K) correspond to possible conditions
in the outer parts of optically thick circumstellar shells.
It follows from Fig. 7 that there is no effect for
$T_{\rm d}({\rm spheroid})/T_{\rm d}({\rm sphere})$,
although the absolute temperature varies
over a fairly wide range: from $T_{\rm d}({\rm sphere})=$ 10.6~{\rm K}
at {\rm $T_{\star}= 500\,$K} to
$T_{\rm d}({\rm sphere})=$ 226.4~{\rm K} at
{\rm $T_{\star}= 10\,000\,$K}.
\begin{figure}\bc
\resizebox{8.5cm}{!}{\includegraphics{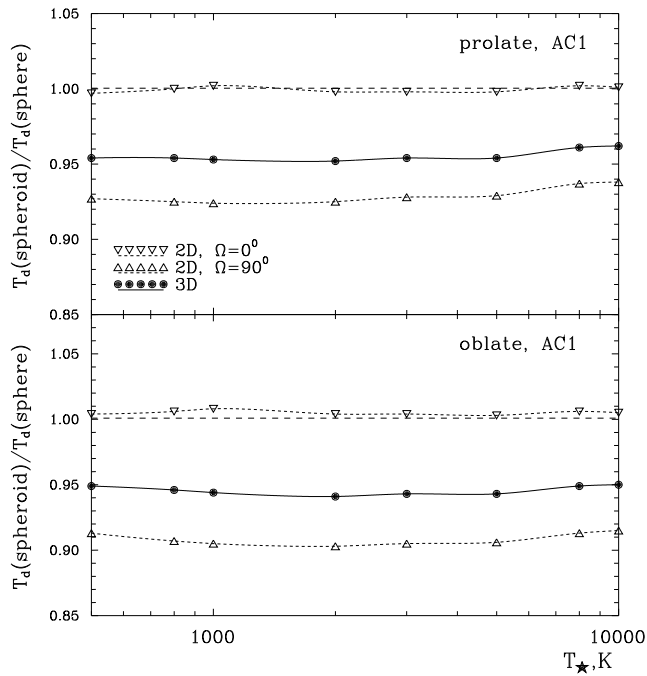}}
\caption[]{
The temperature ratio of prolate and oblate
spheroidal ($a/b = 4$) and spherical amorphous carbon (AC1) grains
with $r_{\rm V}=0.01\,\mkm$ at $R = 10^4~R_{\star}$ from the star.
The sphere temperature is
$T_{\rm d}({\rm sphere})=$ 10.5~{\rm K}, 20.4~{\rm K}, 63.1~{\rm K}, 106.4~{\rm K},
and 226.4~{\rm K} for
 $T_{\star}= 500$~{\rm K}, 1000~{\rm K}, 3000~{\rm K}, 5000~{\rm K},
10\,000~{\rm K}, respectively.
}
\ec\end{figure}

\medskip
\centerline{\it
4.6. Porous Dust Grains and Polarized Incident Radiation
}
\medskip

An increase in the fraction $f$ of vacuum in porous particles causes a
decrease in the effective refractive index, i.e., a reduction in particle
absorptivity. In this case, the shape effects gradually disappear, and
the temperature of spheroidal particles approaches the temperature of spheres.
This effect is illustrated in Table 3, which shows that
$T_{\rm d}({\rm spheroid})/  T_{\rm d}({\rm sphere})$
increases considerably for cellulose
particles at $f \ga 0.5$.

\begin{table*}
\bc
\caption[]{
The temperatures (in K)
of porous spherical and prolate spheroidal ($a/b=10$)
cellulose (cel1000) grains; $r_{\rm V} = 0.01$\,¬ª¬,
$T_{\star}= 2500\,$K, $R = 10\,000\,R_{\star}$}
\smallskip
\begin{tabular}{ccccccc}
\noalign{\smallskip}
\hline
\noalign{\smallskip}
$1-f$  && \multicolumn{3}{c} {$T_{\rm d}(\rm spheroid)/T_{\rm d}(\rm sphere)$}
&& $T_{\rm d}(\rm sphere)$ \\
\noalign{\smallskip}
\cline{3-5}
\noalign{\smallskip}
&& {\rm 2D, $\Omega=0^{\circ}$} & {\rm 2D, $\Omega=90^{\circ}$} & {\rm 3D}&& \\
\noalign{\smallskip}
\hline
\noalign{\smallskip}
1.0 && 0.65 & 0.59 & 0.61  && 59.2 \\
0.9 && 0.65 & 0.59 & 0.61  && 57.7 \\
0.7 && 0.68 & 0.63 & 0.65  && 53.5 \\
0.5 && 0.77 & 0.73 & 0.74  && 47.3 \\
0.3 && 0.94 & 0.91 & 0.92  && 45.0 \\
0.2 && 0.98 & 0.96 & 0.96  && 49.4 \\
0.1 && 0.98 & 0.97 & 0.97  && 54.8 \\
\noalign{ \smallskip}
\hline
\end{tabular}
\ec
\end{table*}

The aligned nonspherical particles in the inner parts of circumstellar
dust shells can polarize the stellar radiation. In principle, the
local linear polarization of scattered radiation can be significant.
The absorption efficiency factors for the TM and TE modes must then
enter into the absorption cross sections [Eqs.~(5), (8), and (9)]
with different weights, depending on the polarization $P$ of the
incident radiation. This causes an increase in the temperature of
prolate spheroids if the electric vector of the incident radiation
is parallel to the particle major axis (in this case,
$Q_{\rm abs}^{\rm TM} > Q_{\rm abs}^{\rm TE}$)
and a decrease in their temperature otherwise
($Q_{\rm abs}^{\rm TM} < Q_{\rm abs}^{\rm TE}$).
The reverse is true for oblate spheroids. However,
the effect is not too large: even at $P = 50\,\%$,
the temperature of prolate spheroidal ($a/b = 10$) cellulose
particles varies between 33.2 and 38.3 K
($T^{\rm 3D}_{\rm d}=$ 36.0~{\rm K} at $P = 0\,\%$).

The following fairly general conclusion can be drawn from what was
considered above: the temperatures of spherical and nonspherical
dust grains are proportional, with the proportionality coefficient
being approximately the same for particles of different sizes at
different distances in the envelopes of stars of diverse
spectral types. This coefficient is determined only by the particle
shape and by absorptive properties of the material of which they are composed.

\bigskip
\centerline{\rm 5. ASTROPHYSICAL IMPLICATIONS}
\medskip

The dust temperature is an important parameter that determines the
infrared spectra of various objects. The efficiency of grain growth and
destruction, formation of molecules on the grain surfaces, and the
alignment of nonspherical particles depend on $T_{\rm d}$ (Voshchinnikov 1986).
Besides, $T_{\rm d}$ determines the polarized submillimeter emission of
nonspherical particles (Onaka 2000) and the gas cooling rate in very
dense interstellar clouds (Whitworth {\it et al.} 1998).

The infrared flux at wavelength $\lambda$ emerging from an optically
thin medium is proportional to the total number $N$ of dust grains
in the medium, the Planck function, which depends on the particle
temperature $T_{\rm d}$, and the emission cross section
$\overline{C}_{\rm em}(\lambda)$ :
\be
F_{\rm IR}(\lambda) = N \ \frac{\overline{C}_{\rm em}(\lambda)}{D^2}
B_{\l}(T_{\rm d})\,,\label{f}
\ee
where $D$ is the distance to the object. The quantities $T_{\rm d}$ and
$\overline{C}_{\rm em}(\lambda)$  in Eq.~(26) depend on the particle shape.

If we assume the chemical composition and sizes of all particles
in the medium to be the same and if we change only the grain shape,
then, when spheres are replaced by nonspherical particles,
the position of the maximum in the spectrum of thermal radiation
is shifted longward, because the temperature of the latter is higher (Fig. 8).
\begin{figure}\bc
\resizebox{8.5cm}{!}{\includegraphics{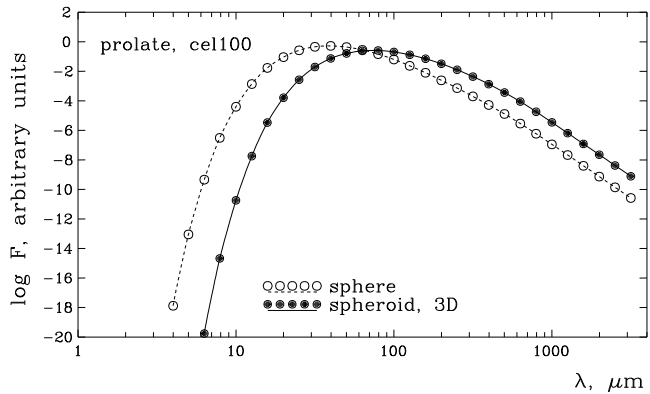}}
\caption[]{
The normalized fluxes emerging from a medium containing
the same (by mass) amount of spherical and prolate spheroidal
($a/b = 10$, 3D orientation) cellulose (cel1000) particles;
$r_{\rm V}=0.01\,\mkm$, $T_{\star}$ = 2500 K, and $R = 10^4~R_{\star}$.
The particle temperatures are $T_{\rm d}({\rm sphere})=$ 59.2~{\rm K}
and $T^{\rm 3D}_{\rm d}=$ 36.0~{\rm K}.
}
\ec\end{figure}

The flux ratio is specified as follows:
\be
\frac{F_{\rm IR}^{\rm spheroid}(\lambda)}{F_{\rm IR}^{\rm sphere}(\lambda)} =
 \ \frac{\overline{C}^{\rm spheroid}_{\rm em}(\lambda)}
        {C^{\rm sphere}_{\rm em}(\lambda)}
 \ \frac{B_{\l}[T_{\rm d}({\rm spheroid})]}
        {B_{\l}[T_{\rm d}({\rm sphere})]} \,.\label{ff}
\ee
It is shown in Fig. 9, where the ratios of the cross sections and the Planck
functions in Eq.~(27) are also displayed. Despite the different
particle temperatures, the Planck functions differ insignificantly.
The main differences in the fluxes result from different emission cross
sections $\overline{C}_{\rm em}(\lambda)$: they are considerably larger for nonspherical
particles at $\lambda > \lambda_{\rm max}$ (Fig. 9). This effect was previously noted by
Ossenkopf and Henning (1994).
\begin{figure}\bc
\resizebox{8.5cm}{!}{\includegraphics{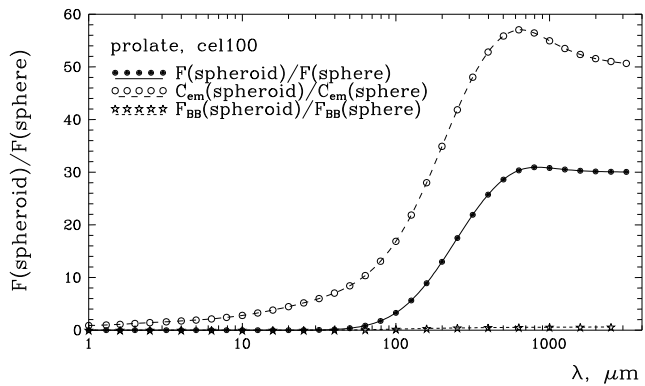}}
\caption[]{
The same as Fig. 8 for the flux ratio. The ratios of the
emission cross sections and fluxes for a blackbody are also shown.
The latter are determined by the particle temperature alone.
}
\ec\end{figure}

If the dust mass in an object is determined from the observed
millimeter flux, then the Rayleigh--Jeans approximation can be used for
the Planck function. For the same flux $F_{\rm IR}(\lambda)$, the dust mass
depends on the particle emission cross sections and temperatures,
while the ratio
\be
\frac{M^{\rm sphere}_{\rm d}}{M^{\rm spheroid}_{\rm d}} =
 \ \frac{\overline{C}^{\rm spheroid}_{\rm em}(\lambda)}
        {C^{\rm sphere}_{\rm em}(\lambda)}
 \ \frac{T_{\rm d}({\rm spheroid})}
        {T_{\rm d}({\rm sphere})}
        \label{mm}
\ee
shows the extent to which the values of $M_{\rm d}$ differ when changing the
grain shape.

The dust mass in galaxies and molecular clouds is commonly
estimated from 1.3-mm observations (Siebenmorgen {\it et al.} 1999). At
this wavelength, the ratio of the cross sections for cellulose particles
with $a/b = 10$ (see Figs. 8 and 9) is $\sim$~50, and the temperature
ratio is $36.0/59.2 = 0.61$, which gives approximately a factor of
30 larger dust mass if the particles are assumed to be spheres rather
than spheroids. In other (not so extreme) cases, the dust-mass
overestimate is much smaller. For example, for prolate amorphous
carbon particles,
$M^{\rm sphere}_{\rm d} / M^{\rm spheroid}_{\rm d} \approx 1.2, 2.3$, and 6.4,
for $a/b =$ 2, 4, and 10, respectively.

The degree of alignment of interstellar and circumstellar dust particles
for rotational (Davies--Green\-stein) orientation depends on the
dust-to-gas temperature ratio: the smaller is this ratio,
the more effective is the alignment of nonspherical particles
(Greenberg 1970; Voshchinnikov 1986). In this case, as follows
from Tables 1 and 2 and from the discussion in subsection 4.4,
the nonspherical particles have a minimum temperature when the
magnetic field is perpendicular to the line of sight ($\Omega = 90\degr$),
i.e., when the polarization of the forward transmitted radiation
is at a maximum. However, this remark more likely refers
to protostellar envelopes than to interstellar particles,
where the dust grains are heated by isotropic radiation.

Yet another important astrophysical process is the formation of
molecules on the grain surfaces. Since the efficiency of this
process strongly depends on the grain temperature, a lower
temperature of nonspherical particles can significantly
facilitate the formation of molecules on their surfaces
[see Voshchinnikov {\it et al.} (1999) for a discussion].

Finally, it should be mentioned that the sublimation
(along with nucleation) of spherical and nonspherical particles
must take place at different distances from the star.
If we consider the evaporation of ice cometary grains in the
interplanetary medium, then, despite the weak shape effect
for ice (see Fig. 6), the differences for spheres and
spheroids prove to be noticeable enough. Ice spheres reach the
sublimation temperature
[$T_{\rm d}({\rm sphere}) = T_{\rm subl}= 100$~{\rm K}]
and begin to
evaporate at a distance $R \approx $~20 AU from the Sun.
For prolate and oblate spheroids ($a/b =$ 10, 3D orientation),
this takes place closer to the Sun: at $R \approx $~18 and 16 AU,
respectively.
    
\bigskip
\centerline{\rm 6. CONCLUSIONS}
\medskip

Nonspherical interstellar and circumstellar dust grains must be slightly
colder than spheres of the same volume composed of the same material.
An analysis of the shape effects for spheroidal particles shows that
the temperature differences do not exceed 10\,\% for a
spheroid aspect ratio {$a/b \approx 2$}. If
$a/b \ge 4$, then the temperature
differences reach 30--40\,\%. Only iron spheroids can be slightly
hotter than spheres. All shape effects are mainly associated
with different emissivities of spherical and nonspherical particles
at low temperatures. Grain porosity causes a reduction in the shape effects,
whereas polarization of the incident radiation can either reduce or
enhance them.

The calculated fluxes at $\lambda \ga 100~\mkm$ turn out to be higher
if the grains are assumed to be nonspherical, resulting in an
overestimation of the dust mass inferred from millimeter observations.

Finkbeiner and Schlegel (1999) point out that the Galactic-dust
thermal emission at long wavelengths gives a major contribution
to the observed background infrared radiation and must be subtracted
in the search for fluctuations of the cosmic microwave background
radiation in its Wien wing.
In this case, very small (in magnitude)
effects compared to which the effect of grain shape is
significant and must be taken into account in modeling are sought.

\newpage
\bigskip
\centerline{\rm ACKNOWLEDGMENTS}
\medskip

We wish to thank V.B. Il'in for the remarks made when reading the
manuscript.
This study was supported by the Volkswagen Foundation,
the Program ``Astronomy" and the grant of INTAS (Open Call 99/652).

\bigskip
\centerline{\rm REFERENCES}
\medskip

\bitem{1. S. Bagnulo, J.G. Doyle, I.P. Griffin,
         Astron. Astrophys. {\bf 301}, 501 (1995).}

\bitem{2. Y. Baron, M. de Muizon, R. Papoular R., B. P$\acute{\rm e}$gouri$\acute{\rm e}$,
         Astron. Astrophys. {\bf 186}, 271 (1987).}

\bitem{3. C. F. Bohren and D. R. Huffman, Absorption and Scattering of
   Light by Small Particles (Wiley, New York, 1983; Mir, Moscow, 1986).}

\bitem{4. B. J. Cadwell, H. Wang, E. D. Feigelson, and M. Frenklach,
   Astrophys. J. {\bf 429}, 285 (1994).}

\bitem{5. A. Z. Dolginov, Yu. N. Gnedin, and N. A. Silant'ev,
   Propagation and Polarization of Radiation in Cosmic
   Medium (Nauka, Moscow, 1979).}

\bitem{6. V.A. Dombrovskii, Doklady AN Armenian SSR {\bf 10}, 199 (1949).}

\bitem{7. B. T. Draine, in Physical Processes in Red Giants,
    Ed. by I. Iben and A. Renzini (Reidel, Dordrecht, 1981), p. 317.}

\bitem{8. H. M. Dyck and M. C. Jennings, Astron. J. {\bf 76}, 431 (1971).}

\bitem{9. Yu. A. Fadeyev, in Circumstellar Matter: IAU Symp.
   No. 122, Ed. by I. Appenzeller and C. Jordan (Reidel,
   Dordrecht, 1987), p. 515.}

\bitem{10. D. P. Finkbeiner and D. J. Schlegel, astro-ph/9907307.}

\bitem{11. A. J. Fleischer, A. Gauger, and E. Sedlmayr, Astron. Astrophys.
    {\bf 266}, 321 (1992).}

\bitem{12. M. E. Fogel and C. M. Leung, Astrophys. J. {\bf 501}, 175 (1998).}

\bitem{13. H.-P. Gail and E. Sedlmayr, Astron. Astrophys. {\bf 132}, 163 (1984).}

\bitem{14. H.-P. Gail and E. Sedlmayr, Astron. Astrophys. {\bf 148}, 183 (1985).}

\bitem{15. J. H. Goebel and H. Moseley, Astrophys. J. Lett. {\bf 290}, L35 (1985).}

\bitem{16. J. M. Greenberg, Interstellar Grains,
 in Stars and stellar systems. Vol.~VII,  Ed.
 by B. M. Middlehurst and L. H. Aller (Univ. Chicago Press, p.~221,
 1968; Mir, Moscow, 1970).}

\bitem{17. J. M. Greenberg, Astron. Astrophys. {\bf 12}, 240 (1971).}

\bitem{18. J. M. Greenberg and G. A. Shah, Astron. Astrophys. {\bf 12}, 250 (1971).}

\bitem{19. J. S. Hall, Science {\bf 109}, 166 (1949).}

\bitem{20. Th. Henning, V. B. Il'in, N. A. Krivova,
B. Michel, N.V. Voshchinnikov, Astron. Astrophys.,
    Suppl. Ser. {\bf 136}, 405 (1999).}

\bitem{21. R. H. Hildebrand, Quart. J. R. A. S. {\bf 24}, 267 (1983).}

\bitem{22. W. A. Hiltner, Science {\bf 109}, 165 (1949).}

\bitem{23. V. B. Il'in and N. V. Voshchinnikov, Astron. Astrophys.,
    Suppl. Ser. {\bf 128}, 187 (1998).}

\bitem{24. C. J\"ager, H. Mutshcke, B. Begemann, J. Dorschner,
Th. Henning,  Astron. Astrophys. {\bf 292}, 641 (1994).}

\bitem{25. C. J\"ager, H. Mutschke, and Th. Henning, Astron. Astrophys.
    {\bf 332}, 291 (1998).}

\bitem{26. M. Jura, Astrophys. J. {\bf 434}, 713 (1994).}

\bitem{27. M. Jura, Astrophys. J. {\bf 472}, 806 (1996).}

\bitem{28. J.-P. J. Lafon and N. Berruyer, Astron. Astrophys. Rev. {\bf 2}, 249 (1991).}

\bitem{29. P. L. Lamy and J.-M. Perrin, Astron. Astrophys. {\bf 327}, 1147 (1997).}

\bitem{30. I. Little-Marenin, Astrophys. J. Lett. {\bf 307}, L15 (1986).}

\bitem{31. S. Lorenz-Martins and J. Lef$\acute{\rm e}$vre, Astron. Astrophys. {\bf 291}, 831 (1994).}

\bitem{32. J. S. Mathis, P. G. Mezger, and N. Panagia, Astron. Astrophys.
    {\bf 128}, 212 (1983).}

\bitem{33. T. Onaka, Astrophys. J. {\bf 533}, 298 (2000).}

\bitem{34. V. Ossenkopf and Th. Henning, Astron. Astrophys. {\bf 291}, 943 (1994).}

\bitem{35. V. Ossenkopf, Th. Henning, and J. S. Mathis, Astron. Astrophys.
    {\bf 261}, 567 (1992).}

\bitem{36. B. P$\acute{\rm e}$gouri$\acute{\rm e}$, Astrophys. Space Sci. {\bf 136}, 133 (1987).}

\bitem{37. S. J. Shawl, Astron. J. {\bf 80}, 602 (1975).}

\bitem{38. R. Siebenmorgen, E. Kr\"ugel, and R. Chini, Astron.
    Astrophys. {\bf 351}, 495 (1999).}

\bitem{39. L. Spitzer, Jr., Physical Processes in Interstellar
    Medium (Wiley, New York, 1978; Mir, Moscow, 1981).}

\bitem{40. H. C. van de Hulst, Rech. Astron. Obs. Utrecht {\bf 11}, Part 2 (1949).}

\bitem{41. N. V. Voshchinnikov, Itogi Nauki Tekh., Ser. Issled. Kosm.
    Prostranstva {\bf 25}, 98 (1986).}

\bitem{42. N. V. Voshchinnikov and V. G. Farafonov, Astrophys.
    Space Sci. {\bf 204}, 19 (1993).}

\bitem{43. N. V. Voshchinnikov, D. A. Semenov, and Th. Henning,
    Astron. Astrophys. {\bf 349}, L25 (1999).}

\bitem{44. N. V. Voshchinnikov, V. B. Il'in, Th. Henning,
    B. Michel, V.G. Farafonov,  J. Quant. Spectrosc. Radiat. Transf. {\bf 65}, 877 (2000).}

\bitem{45. L. B. F. M. Waters, F. J. Molster, and C. Waelkens,
    in Solid Interstellar Matter: the ISO Revolution,
     Ed. by L. d'Hendecourt {\it et al.} (Springer-Verlag, Berlin, 1999), p. 219.}

\bitem{46. D. C. B. Whittet, Dust in the Galactic Environments
    (Institute of Physics Publ., New York, 1992).}

\bitem{47. A. P. Whitworth, H. M. J. Boffin, and N. Francis,
    Mon. Not. R. Astron. Soc. {\bf 299}, 554 (1998).}

\bitem{48. P. Woitke, C. Dominik, and F. Sedlmayr, Astron. Astrophys.
    {\bf 274}, 451 (1993).}

\bigskip
\bigskip

\rightline {\it Translated by V. Astakhov}

\end{document}